# Why every observatory needs a disco ball

Robert J. Cumming[1], Alexander G.M. Pietrow[2], Livia Pietrow[3], Maria Cavallius[4], Dominique Petit dit de la Roche[5], Casper Pietrow[3], Ilane Schroetter[6], Moa Skan[7]

1 Onsala Space Observatory, Chalmers University of Technology, SE-439 92 Onsala, Sweden

2 Leibniz-Institut für Astrophysik Potsdam (AIP), An der Sternwarte 16, 14482 Potsdam, Germany

3 Leiden University, P.O. Box 9513, 2300 RA Leiden, the Netherlands

4 Department of Astronomy, Stockholm University, 106 91 Stockholm, Sweden

5 Observatoire de l'Université de Genève, Chemin Pegasi 51, 1290 Versoix, Switzerland

6 Institut de Recherche en Astrophysique et Planétologie (IRAP), Université Toulouse III - Paul Sabatier, CNRS, CNES, 9 Av. du colonel Roche, 31028 Toulouse Cedex 04, France

7 Institute for Solar Physics, Department of Astronomy, Stockholm University, Roslagstullsbacken 21, Stockholm, Sweden

**Email**:
robert.cumming@chalmers.se, apietrow@aip.de

**Abstract**

Commercial disco balls provide a safe, effective and instructive way of observing the Sun. We explore the optics of solar projections with disco balls, and find that while sunspot observations are challenging, the solar disk and its changes during eclipses are easy and fun to observe. We explore the disco ball's potential for observing the moon and other bright astronomical phenomena.

## 1. Introduction

Disco balls, also commonly known as mirror balls, have the potential to provide a safe and engaging way of viewing the solar disk. In this paper, we give an introduction to using disco balls for astronomical observations, with a focus on their potential for public engagement and education. We pay special attention to using disco balls for viewing (partial) solar eclipses (See Fig. 1). We compare the disco ball's advantages to other eclipse viewing methods, in terms of safety, accessibility and entertainment value.

For astronomical purposes, the optics of disco balls are similar to the optics of the better-known pinhole camera. Disco balls are collections of what are known as pinhead mirrors, each the reflective equivalent of a pinhole camera aperture. The pinhole camera, a simple yet elegant optical device, serves as an invaluable educational tool for exploring the fundamentals of optics while offering numerous practical applications (Young 1989). In astronomy the pinhole camera is most widely used for solar projections, serving as an inexpensive and safe way to observe the solar eclipses or to observe sunspots. In the literature, the pinhead mirror has been rediscovered on multiple occasions, but only rarely treated as an astronomical instrument. While Takeda (2014) attributes their discovery to Wood (1934), pinhead mirrors were described in detail by Nilsson (1986) and later patented (1989; expired 2007; patent nr. US4948211A). We have found experiments imaging the sun with pinhole mirrors described on the web by Malpas (Pinhole astrophotography, http://users.erols.com/njastro/barry/) and by Takeda (Laboratory: Pinhole photography, http://atelier.bonryu.com/en/welcome/lensless/phphoto-l/phvariation_1_phmirror/).

While pinhead mirrors are relatively accessible by means of breaking or covering larger mirrors, a safe alternative has become more affordable in recent years. Disco balls, consisting of hundreds of small mirror segments, are readily available at remarkably low prices. Additionally, their associations to concerts, discotheques, and parties make them interesting and unexpected objects for demonstrating physics to schoolchildren and the general public.

In this paper, we expand upon the traditional applications of disco balls. By exploring their ability to safely and engagingly image the Sun, we explore how effectively disco balls can bridge the gap between complex astronomical concepts and the public's understanding, evaluating them as a resource for educators, outreach coordinators, and astronomy enthusiasts (see Prados Ribeiro (2016) for a complementary approach). We are particularly motivated by the solar eclipse of April 2024, which is expected to attract large public interest and potentially many millions of partial eclipse observers (Duncan 2023).

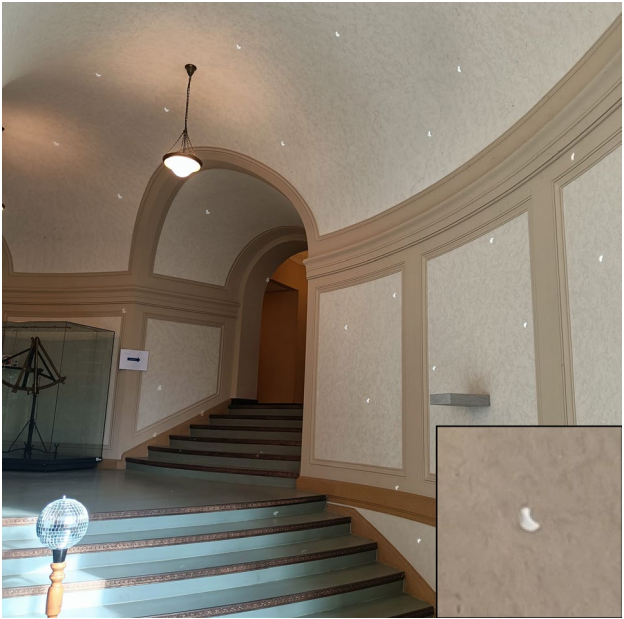

Fig. 1. A disco ball projector during the partial solar eclipse of 25 October 2022 in Potsdam, Germany. An enlarged solar image is shown in the lower right corner.

## 2. Method

In this section we describe the optical principles of pinhole projectors, and how these can be applied to pinhead mirrors.

### 2.1 The pinhead mirror and its focus

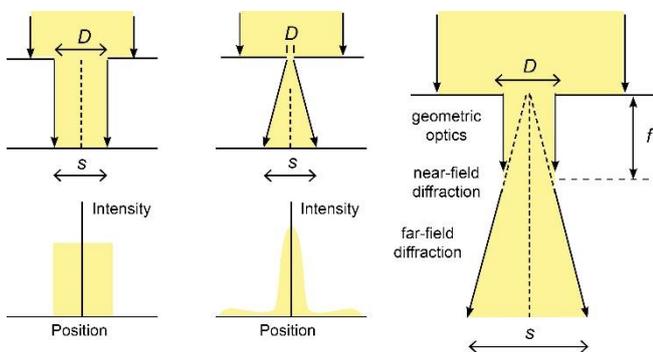

Fig. 2. Pinhole optics. Left: A large pinhole that projects a geometrical shadow. Middle: A small pinhole that projects a diffraction pattern. Right: Both regimes combined with the near field diffraction between the two regimes. After Young (1989).

As with lenses and mirrors, the physics of pinholes and pinhead mirrors are largely the same, with the biggest difference being that lenses and pinholes are transmissive devices, while pinhead mirrors are reflective. For this reason we briefly discuss the physics of pinhole projectors and then discuss how they apply to pinhead mirrors.

Pinhole projectors can be described in much the same way as lenses, with a diameter ($D$) and a focus distance ($f$). When the pinhole is large then the image created by it will simply be the geometrical shadow of the pinhole (see Fig. 2a), with its diameter ($s$) being equal to the pinhole diameter D. This is also called the pupil plane. When the pinhole is very small, then instead you will see a diffraction pattern (see Fig. 2b) which is given by $s = 2.44\lambda f/D$ for a round pinhole and $s = 2\lambda f/D$ for a square one. Here, $s$ is the diameter of the spot, and $\lambda$ the wavelength of the observed light. This is typically taken to be the middle of the visual spectrum at 550 nm.

Following Young (1971, 1989) it can be shown that the ideal focus of a pinhole lies between these two limits, in the so-called near-field diffraction regime (see Fig. 2c). For a square pinhole this is given by

$$f = D^2/2\lambda \qquad (1$$

A typical inexpensive disco ball has mirror segments that range in widths between 0.4 and 1 cm. (The small mirrors are typically found on smaller disco balls sold as Christmas tree decorations or sometimes even earrings. High-end professional disco balls tend to also have small mirrors of around 0.5 cm that are much more densely packed.)

Filling in the aforementioned mirror widths, and 550 nm for the wavelength into equation 1, gives us the following relation,

$$f = 9091\, D^2, \qquad (2$$

with $f$ and $D$ both in cm. We find an optimal focus of about 1500 cm and 9100 cm respectively, showing that even the smallest mirrors require large distances to focus completely.

However, as anyone who has observed with a telescope before can attest, perfect focus isn't necessary in order to discern the round shape of a planet and even some of its details (The defocus can also be expressed mathematically, as is shown by Baird (1980)). We observed the Sun with different disco balls on different occasions in order to explore how image quality affects ability to observe different phenomena, in particular eclipse phases and sunspots.

Solar observations can be used to illustrate the difference between the near field pupil plane and the mid field focal plane. The projected images turn from squares to circles as one moves away from the disco ball. This property can be used to

Cumming *et al* (submitted to Physics Education)

show why stars imaged by modern telescopes can be round despite central obscurations on their mirrors.

The diameter of the solar image (*d*) can also be expressed with simple geometry (e.g. Kriss 1996), since we know that the Sun spans roughly half a degree on the sky, resulting in

$$d = f/108 \qquad (3$$

This results in a spot of roughly 9 cm diameter at 10 m distance for the large segments.

## 2.2 Contrast and projection quality

It is possible to estimate the brightness of the projected image by using the source's intensity in lux, which is 120 000 for direct sunlight. The illuminance, $I_0$, given in lumen/m$^2$ (or lux) can then be multiplied by the area of the pinhole ($A_p$) to get the luminance of the pinhole. This can then be divided by the projected area of the spot ($A_s$) to get the illuminance of the projected spot,

$$I = I_0 A_p / A_s. \qquad (4$$

This equation assumes that 100% of the light gets reflected from the surface on which it is projected. In practice this will depend on the surface. A white sheet of smooth paper will offer the best results.

When letting sunlight through a 1 cm$^2$ pinhole, this becomes 12 lumen. If focused into a solar disk with a 10 cm diameter, the resulting illuminance is roughly 1500 lux, while a 5 cm spot will have an illuminance of roughly 6100 lux.

A well-lit classroom has around 300 to 500 lux, an average living space close to 50, a darkened room 10 or below, and the outdoors can easily be above 1000 in the shade alone (e.g. Bhandary 2021). This means that the contrast of the image will depend on where the disco ball is placed, and can be calculated by simply dividing the two numbers. If a picture is taken of the projected disk, it can also be calculated by finding the average background and the average disk intensity. Regardless, a small (~5 cm) spot should be easily visible in most environments.

Sources other than the Sun can also be projected using a pinhole mirror, but given that these are typically much less bright, one has to do this in a darkened room. A light bulb is a good target, as it will project the shape of the bulb if viewed far enough away. The Moon is difficult as at its brightest it is around 400 000 times dimmer than the Sun, with a source brightness of just 0.32 lux (Kyba 2017). This means that a 5 cm spot cast by a 1 cm$^2$ mirror would have an illuminance of $1.63\times10^{-4}$ lux, which is at the limit of the human eye, though within reach of a camera.

For some mirror balls we have tested, overlapping ghost images limit the image quality but reveal interesting mirror optics. In such cases a secondary image appears on either side of the primary (see Fig. 3), likely caused by reflections between the two surfaces of the mirror. The ghost images resemble the effect caused by a diffraction pattern, but we judge that diffraction effects are not detectable for disco ball images at reasonable distances. If the front and back surfaces of each mirror segment are not parallel then the resulting wedge can produce multiple images from the incident beam (Beers 1974). In what way and how much the images are dispersed depends on the angle of incidence and the orientation of the mirror plane. If the incident beam meets a wedge, the images may be dispersed by up to an angle of 4*nA*, where n is the refractive index of the mirror glass and *A* is the deviation from planarity in radians (Beers 1974). The multiple images we measure correspond to values of *A* of about a hundredth of a degree.

If we assume that the front surface has a reflectivity of 4%, then the first ghost will be of that strength. Then the central image will have an intensity of $0.96 \times 0.96 = 92\%$, and the second ghost of $0.96 \times 0.04 \times 0.96 = 3.7\%$. Further ghosts will have intensities of less than 0.1%, and likely not be visible. This is why professional astronomical mirrors are usually front-coated. This effect can be further demonstrated by shining a laser on the segments, which should also produce 3 spots.

Not all disco balls suffer from this effect, but we have seen it in both expensive and cheap models, perhaps due to a single sheet of imperfect glass having been broken up into the segments used to coat the entire ball. In such a case we recommend buying a new ball from a different source, as to minimise the chance of getting one from the same batch.

A student should be able to calculate the reflectivity of the mirror by comparing the brightness of the three spots and solving for the equations given above.

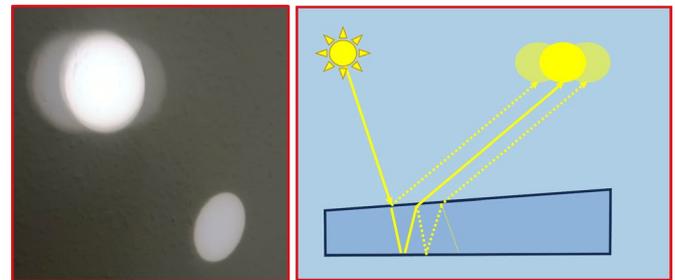

Fig. 3. Ghost images are produced by back-coated mirrors. Left: A photograph showing a solar image from a disco ball with a ghosting effect, and one without. The latter has a much sharper disk. Right: A diagram explaining the ghosts.

## 3. Results

In this section we assess how a disco ball projecting the Sun and other sources can be used as an educational tool. We report preliminary results from testing disco ball observations with the public.

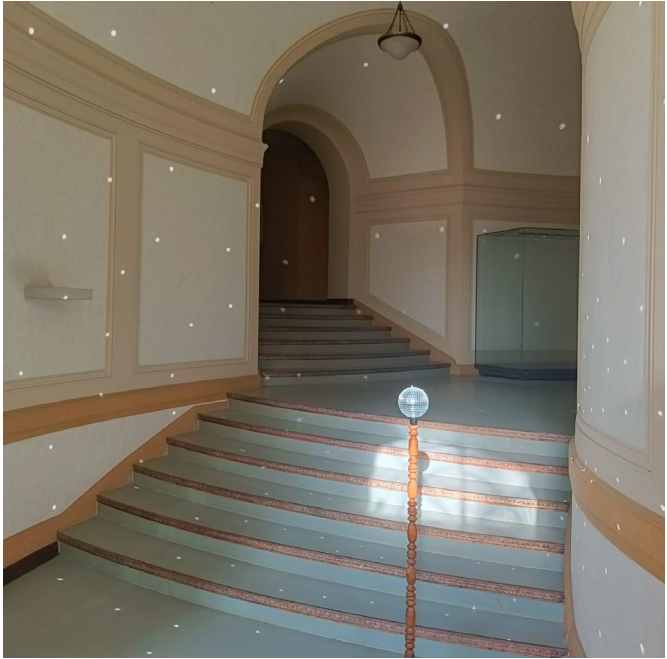

Fig 4: A disco ball safely projecting the Sun all over a room.

### 3.1 Observing the solar disk and eclipses

Pinholes have been used to observe the Sun since antiquity (see for example Heilbron 1999; Sigismondi & Fraschetti 2001). Unlike commonly used tools for projecting eclipses, such as pinhole projectors, colanders, or tree canopies, a disco ball is able to function for large crowds because it does not merely work on the area where it casts its shadow, but across the entire illuminated hemisphere, which can project solar images across an entire room or courtyard (see Fig. 4). Additionally, it does not need to be pointed or moved to project the Sun, as new segments get illuminated as the Sun moves out of the older ones. This means that it is enough to simply place a disco ball close to a window in order to fill a large part of the space with solar projections. In fact, the disco ball encourages a crowd to disperse, as they walk towards the walls to look at the projected images.

The disco ball is also safe in the sense that its images diverge with distance, making looking into the ball comparable to the glare that one sees when the reflection of shiny objects enters one's eyes. Even if someone comes close to the mirrors where the reflection is more concentrated, their head would block out the reflection and draw the attention of the demonstrator.

Our test observations of the partial solar eclipse visible in northern Europe on 25 October 2022 were successful. We observed a clear crescent shape that changed as the eclipse progressed (see Fig. 1). Similar eclipse observations have been reported on the web (J. Hamlin, 2012, "Disco ball projects solar eclipse all over back yard", https://youtu.be/6dUcQ2JAJd4, annular solar eclipse of May 20, 2012).

We have also observed leaves and larger obscurations between the Sun and the disco ball (See Figs. 5a and 5b) that become more or less focused depending on how far they are from the mirrors.

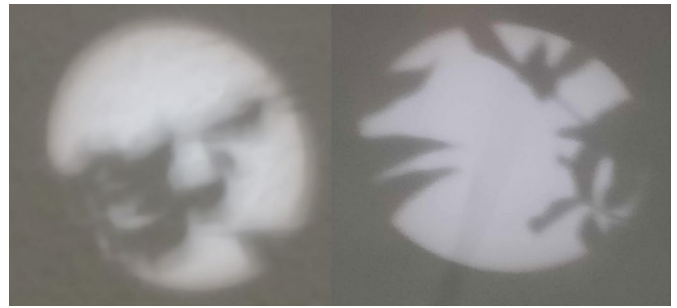

Fig 5a: Solar projections with obscurations. Left: Out of focus leaves close to the disco ball. Right: In-focus leaves situated further away from the disco ball.

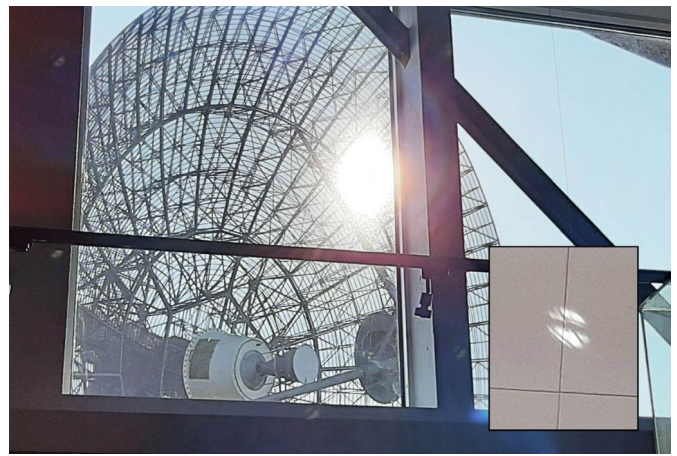

Fig 5b: Solar projections with obscurations. This view, taken from close to the disco ball, shows images on the wall to the left of the window obscured by the structure of a radio telescope dish. An enlarged solar image is shown in the lower right corner.

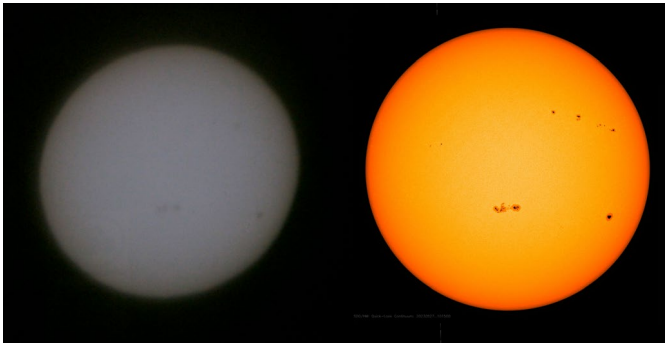

Fig 6. A sunspot group observed on 2023 May 27 observed with a disco ball projector (left) and observed with the SDO/HMI Satellite (right; Scherrer 2011).

*3.2 Observing sunspots and transits*

We have successfully used disco balls to observe sunspots. From around 6 m, sunspots are easy to discern when projected on walls or sheets of paper. To maximise contrast, we have found it is best to project onto high-quality stiff white card rather than directly on walls or on ordinary paper.

For the disco balls we tested during 2022 and 2023, a period of moderate to strong solar activity, only the largest sunspots are visible. We estimate a lower size limit of 1-2 arcminutes (see Fig. 6), which indicates that large sunspots can regularly be observed during around half of each solar cycle.

Based on our sunspot measurements, we are confident that the next transit of Venus in 2117 will be observable with a disco ball. A pinhead mirror was used to image the 2012 transit and reported on the web (C. Barry, 2012, http://341ontheriver.com/visions-of-venus/). Transits of Mercury are not likely to be detectable due to the small (~0.2 arcminute) angular size of the planet.

*3.3 Observing the Moon and other objects*

Besides the Sun, the disco ball can also be used to project other natural light sources, as long as they are bright enough and the projection environment allows sufficient contrast.

We tried on a number of occasions to observe different phases of the Moon using disco balls. We were successful with the full moon on 2022 December 7, but could only see near-field images. Success is dependent on projecting towards a dark surface and without other light sources nearby. We expect that far-field images should be detectable using a camera.

*3.4 Exploratory user testing*

During the period March-May 2023 we tested a portable disco ball as part of a permanent exhibition at a university observatory, receiving visitors in groups of all ages. When illuminated, the ball was popular with visitors. In particular children enjoyed the opportunity of spinning the ball and watching the reflected images move across the walls.

We have attempted to assess qualitatively how much support visitors might need to use the disco ball as a way of exploring physics and astronomy on their own. We used a neutral approach, asking questions and being careful not to oversell the disco ball's potential. We asked groups of three to five schoolkids (age around 15) to think about how the ball works, asking "Why do you think there is a disco ball in an exhibition about space and astronomy?" Neither group mentioned the possibility of observing the Sun. One student guessed that the ball itself symbolised a star. Another thought that the images on the walls and ceiling were somehow light from stars. We asked these and other young visitors why they think the images are round. None of the students suggested that the roundness was due to the shape of the Sun on the sky.

During an eclipse, other test subjects easily recognised the partially covered solar disk. We observed a partial solar eclipse with a disco ball during an observatory open day on 25 October 2022. On this occasion, visitors we interviewed clearly understood that they were looking at the Sun because they saw the crescent shape characteristic of the eclipse (Fig. 1). During a public event in May 2023 we noted that while the possibility of observing the very prominent sunspot group AR 13315 using the disco ball excited response from the audience to a talk by one of us, discerning poorly visible dark patches in solar images was less popular than other visitor-oriented content.

We aim to test this method of solar projection on a larger scale in the near future in conjunction with the April 2024 solar eclipse.

## 4. Conclusions

Disco balls can be used with minimal preparation as mildly defocused pinhole projectors that are able to project bright light sources (such as the Sun) on a wall or screen. The disco ball serves as an intuitive tool for explaining the pupil and focal plane. While light bulbs and other artificial sources can be used to explore the optics of pinholes, we have focused on exploring astronomical objects and in particular the Sun.

Unlike more traditional solar projection tools like pinhole projectors and colanders, the disco ball spreads its solar images across a room, producing recognizable solar disks from distances of about 2 metres and onwards. This makes the disco ball a more accessible tool for larger or socially distanced groups. It is also possible to observe large sunspots with a disco ball with small enough mirror segments. We expect that observing the Moon and its phases with a disco ball is possible, but requires a darker environment than we have yet been able to achieve.

We suggest that many solar observation experiments for small telescopes or pinhole cameras can also be carried out using a disco ball projector. The advantage is that several students could work on their own spots in different parts of the room, thus allowing them to later compare their observations. For example, one could calculate the speed at which the Earth rotates around its axis by measuring the movement of the projected spots (Prados Ribeiro 2016), measure the change of the intensity over time during an eclipse by photographing the projected spots (Slaton 2021; Bensky 2022), determine the ellipticity of the Earth's orbit by comparing seasonal changes in the solar disk size (Wootton 2001), measure the solar limb darkening (Inbanathan 2021), and even study active regions if they are large enough (Skan et al. 2023).

We believe that the disco ball is a versatile and engaging tool for educational purposes, deserving wider use both for classroom demonstrations and for public events.

## Acknowledgements

AP is supported by grants from the European Commission's Horizon 2020 Program under grant agreements 824064 (ESCAPE – European Science Cluster of Astronomy & Particle Physics ESFRI Research Infrastructures) and 824135 (SOLARNET – Integrating High-Resolution Solar Physics).

The contributions of DP have been carried out within the framework of the NCCR PlanetS supported by the Swiss National Science Foundation under grants 51NF40_182901 and 51NF40_205606.

This research has made use of NASA's Astrophysics Data System (ADS) bibliographic services.

We thank Gilles Otten for discussions. We thank David Bailey and the r/askphysics community for the stimulating discussion on pinhole optics. We thank Kelley Hess, Ola Schönbeck and José Sánchez for discussions and invaluable help with experimental equipment.

## References

Baird, L. C. (1980). How big is a pinhole? In Medical Physics (Vol. 7, Issue 1, pp. 64–64). Wiley. https://doi.org/10.1118/1.594660

Beers, Y., 1974. The theory of the optical wedge beam splitter (Vol. 146). National Bureau of Standards https://nvlpubs.nist.gov/nistpubs/Legacy/MONO/nbsmonograph146.pdf

Bensky, T. (2022). Sun Photometry for Introductory Students. In The Physics Teacher (Vol. 60, Issue 9, pp. 728–731). American Association of Physics Teachers (AAPT). https://doi.org/10.1119/5.0062948

Bhandary, S. K., Dhakal, R., Sanghavi, V., & Verkicharla, P. K. (2021). Ambient light level varies with different locations and environmental conditions: Potential to impact myopia. In I.-J. Wang (Ed.), PLOS ONE (Vol. 16, Issue 7, p. e0254027). Public Library of Science (PLoS). https://doi.org/10.1371/journal.pone.0254027

Duncan, D. (2021) Prepare for the 2023 and 2024 Solar Eclipses! School and Community Events and Fundraising. The Physics Teacher 1 May 2023; 61 (5): 334–338. https://doi.org/10.1119/5.0131185

Gomez, E., 2013, Measure the Solar Diameter, astroEDU, 1305, doi:10.11588/astroedu.2013.1.81186, https://astroedu.iau.org/en/activities/measure-the-solar-diameter/

Heilbron, J.L., 1999. The sun in the church. The Sciences, 39(5)

Hewitt, P. (2000). Pinhole image of the Sun. In The Physics Teacher (Vol. 38, Issue 5, pp. 272–272). American Association of Physics Teachers (AAPT). https://doi.org/10.1119/1.1528659

Inbanathan, S. S. R., Moorthy, K., \& S, A. K. (2021). Observing Solar Limb Darkening in the Classroom. In The Physics Teacher (Vol. 59, Issue 4, pp. 292–293). American Association of Physics Teachers (AAPT). https://doi.org/10.1119/10.0004162

Kriss, V. (1996). Measuring pinhole images of the sun. In The Physics Teacher (Vol. 34, Issue 3, pp. 190–191). American Association of Physics Teachers (AAPT). https://doi.org/10.1119/1.2344399

Kyba, C. C. M., Mohar, A., & Posch, T. (2017). How bright is moonlight? In Astronomy & Geophysics (Vol. 58, Issue 1, p. 1.31-1.32). Oxford University Press (OUP). https://doi.org/10.1093/astrogeo/atx025

Nilsson, T. H, 1986, Applied Optics, DOI: 10.1364/AO.25.002863

O'Shea, D. 1987, Lasers and Optronics, July 1987, p. 12

Prados Ribeiro, J. L., 2016, Reflections on a Disco Ball, The Physics Teacher 54, 532–534 https://doi.org/10.1119/1.4967890

Schaefer, B. E. 1993, ApJ, 411, 909-919

Scherrer, P. H., Schou, J., Bush, R. I., Kosovichev, A. G., Bogart, R. S., Hoeksema, J. T., Liu, Y., Duvall, T. L., Jr., Zhao, J., Title, A. M., Schrijver, C. J., Tarbell, T. D., & Tomczyk, S. (2011). The Helioseismic and Magnetic Imager (HMI) Investigation for the Solar Dynamics Observatory (SDO). In Solar Physics (Vol. 275, Issues 1–2, pp. 207–227). Springer Science and Business Media LLC. https://doi.org/10.1007/s11207-011-9834-2

Sigismondi, C. and Fraschetti, F., 2001. Measurements of the solar diameter in Kepler's time. The Observatory, Vol. 121, p. 380-385

Skan, M., Pietrow, A.G.M., Åkerman, H.,  Kiselman, D., (Submitted), Measuring the temperature of sunspots with state-of-the-art Solar observations.

Slaton, W. V., Jeffery, E. (2021). Balloon-borne Solar Radiation Measurements During the 2017 North American Eclipse. In The Physics Teacher (Vol. 59, Issue 5, pp. 328–332). American Association of Physics Teachers (AAPT). https://doi.org/10.1119/10.0004880

Takeda, T. (2014), Pinhole Photo Encyclopedia, NextPublishing Authors Press

Wood, R. (1934), Physical Optics, 3rd edition, https://archive.org/details/physicaloptics03woodgoog

Wootton, A. (2001). Size of the Sun, The Physics Teacher (Vol. 39, Issue 4, pp. 249–250). American Association of Physics Teachers (AAPT). https://doi.org/10.1119/1.1367799

Young, M. 1971. Pinhole Optics. In Applied Optics (Vol. 10, Issue 12, p. 2763). The Optical Society. https://doi.org/10.1364/ao.10.002763

Young, M. 1989, The pinhole camera: Imaging without lenses or mirrors, The Physics Teacher 27, 648 https://doi.org/10.1119/1.2342908